\documentclass[doublecol,linenumbers]{epl2}
\usepackage{amsmath}
\usepackage{graphicx}

\title{Diffusion in a system of a few distinguishable fermions in a one-dimensional double-well potential}

\author{Tomasz Sowi\'nski\inst{1} \and Mariusz Gajda\inst{1} \and Kazimierz Rz\c a\.zewski\inst{2}}
\shortauthor{T. Sowi\'nski \etal}

\institute{                     
  \inst{1} Institute of Physics of the Polish Academy of Sciences, Al. Lotnik\'ow 32/46, 02-668 Warszawa, Poland\\
  \inst{2} Center for Theoretical Physics of the Polish Academy of Sciences, Al. Lotnik\'ow 32/46, 02-668 Warszawa, Poland  
  }

\abstract{
Dynamical properties of a few ultra-cold fermions confined in a double-well potential is studied.  We show that the dynamics, which is governed by single-particle tunnelings for vanishing interactions, is completely different for strong interactions. Depending on the details of the configuration, for sufficiently strong interactions (repulsions or attractions) the particle flow through the barrier can be accelerated or slowed down. This effect cannot be explained with the single-particle picture. It is clarified with direct inspection to the spectrum of the few-body Hamiltonian. 
}

\begin{document}

\maketitle 

\section{Introduction}
Due to the remarkable progress in controlling and engineering systems of ultra-cold gases, the famous concept of the quantum simulator proposed by Feynman over 30 years ago \cite{Feynman} is undergoing amazing renaissance. Many old-fashioned toy-models known from classical textbooks on quantum theory are realized in nowadays experiments and their properties can be probed with very high accuracy \cite{BlochReview}. The famous work-horses of statistical and condensed matter physics like Ising, Heisenberg, or Hubbard models, together with plethora of their variations, are currently accessible in many laboratories in many different configurations \cite{LewensteinBook, DuttaReview}. These experimental possibilities offer us a direct insight to the structure of strongly-correlated states of the system not only in statical but also in dynamical situations. One of the prominent examples is the in situ observation of the system undergoing the quantum phase transition \cite{Greiner1,Bakr,Chen}.

Completely new brand of experiments are performed in the Johim Selim's group, where ultra-cold mixtures of a few fermions loaded in an effective one-dimensional trap are realized \cite{Serwane,Wenz}. The shape of trapping potential together with the strength of the interaction between particles of different spins can be tuned dynamically and occupations of the single-particle levels can be measured. This technique enables one to observe the formation of the Fermi sea when consecutive particles are added to the system one by one \cite{Wenz}. In other experiments the fermionization of distinguishable particles \cite{Zurn1,Busch1998,SowinskiGrass} as well as pairing for attractive forces \cite{Zurn2,SowinskiPairs,Amico,SowinskiPairs2} were observed. Recently, many theoretical results on a few ultra-cold fermions in harmonic confinement were also reported \cite{Blume1,Blume2,Gharashi,Bugnion,Brouzos,Deuretzbacher,Cui,Lindgren,Pecak} and they are awaiting experimental verification.

In the most recent experiment \cite{Murmann} a full control of two fermions in a double-well potential was achieved. This achievement, in the light of double-well experiments with many fermions or bosons \cite{Catoliotti,Shin,Albiez,Levy,LeBlanc,Milburn,XinWei,Valtolina}, have opened new possibility of studying the celebrated Josephson effect \cite{Josephson} in completely new manner, i.e. from the few-body perspective. It is expected that upcoming experiments will give some answers to still opened questions on the dynamical properties of a few quantum particles \cite{BlumeReview,Zinner2014}. 

Motivated by these beautiful experiments, here we analyze dynamical properties of the two-flavored mixture of a few fermions confined in a one-dimensional double-well potential. We consider a scenario with initially separated clouds of opposite fermions in distant traps which are brought nearby instantaneously. This initial state corresponds to two magnetic domains of the famous Stoner model \cite{Stoner} of itinerant ferromagnetism. However, it should emphasized that this analogy originates in spatial separation rather than in interactions. Although, in the context of cold-atoms, predictions of the model were qualitatively confirmed \cite{Keterle}, some open questions are still under debate and previous conclusions were questioned \cite{Sanner,Sanner2,Pekker}. From the few-body perspective it is also believed that some precursors of the Stoner domains can be experimentally observed \cite{Zinner2014,Pecak}. The whole analysis is performed with numerically direct and exact method originating in the exact diagonalization of the few-body Hamiltonian. 
 
\section{The model}

We consider an ultra-cold system of fermions of mass $m$ confined in a one-dimensional double-well potential and interacting via short-range interactions. In the second quantization the Hamiltonian has a form:
\begin{multline} \label{Ham1}
\hat{\cal H} = \sum_\sigma\int\!\!\mathrm{d}x\,\hat{\boldsymbol{\Psi}}_\sigma^\dagger(x) {\cal H}_0\hat{\boldsymbol{\Psi}}_\sigma(x) \\
+g\int\!\!\mathrm{d}x\,\hat{\boldsymbol{\Psi}}_\downarrow^\dagger(x)\hat{\boldsymbol{\Psi}}_\uparrow^\dagger(x)\hat{\boldsymbol{\Psi}}_\uparrow(x)\hat{\boldsymbol{\Psi}}_\downarrow(x),
\end{multline}
where ${\cal H}_0=-\frac{\hbar^2}{2m}\frac{\mathrm{d}^2}{\mathrm{d}x^2}+V(x)$ is the single-particle Hamiltonian with an external potential $V(x)$. The field operator $\hat{\boldsymbol{\Psi}}_\sigma(x)$ annihilates a fermion with spin $\sigma\in\{\uparrow,\downarrow\}$ at point $x$. Fermionic field operators of a given spin obey standard anti-commutation relations $\left\{\hat{\boldsymbol{\Psi}}_\sigma(x),\hat{\boldsymbol{\Psi}}_\sigma(x')\right\}=0$ and $\left\{\hat{\boldsymbol{\Psi}}_\sigma(x),\hat{\boldsymbol{\Psi}}^\dagger_\sigma(x')\right\}=\delta(x-x')$. We want to point out that for the case studied here, dynamical equations generated by the Hamiltonian \eqref{Ham1} do not depend on the commutation
relations assumed between the different species atoms. Typically, as both kinds of atoms belong to the fermion family, one is tempted to assume anti-commutation relation between them. Note however, that our Hamiltonian does not transform one kind of atoms into the other kind, i.e. is a biquadratic form in field operators and the same dynamical equations will be obtained if commutation rules are assumed instead (even number of commutation  operations is required). The physical reason  for this fact is that in the model studied, both species are fundamentally distinguishable \cite{WeinbergBook}. In such a case the wave function does not have to be neither symmetric nor antisymmetric with respect to exchange of two particles of the different kind. It is quite obvious that the Hamiltonian (\ref{Ham1}) commutes with the total number of fermions of a given spin $\hat{\cal N}_\sigma=\int\mathrm{d}x\,\hat{\boldsymbol{\Psi}}^\dagger_\sigma(x)\hat{\boldsymbol{\Psi}}_\sigma(x)$.

For convenience we model the double-well potential by the function
\begin{figure}
\centering
 \includegraphics[scale=1.05]{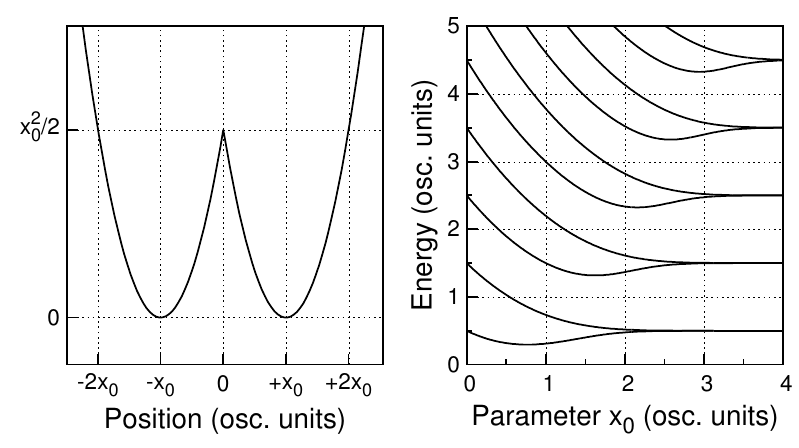}
 \caption{Properties of the external potential used to model the double-well confinement. The parameter $x_0$ simultaneously controls the distance between wells and the barrier between them. Note, that independently of $x_0$, each well in the vicinity of its minimum can be approximated by the harmonic traps with fixed frequency $\Omega$. (right panel) Eigenergies of the single-particle Hamiltonian ${\cal H}_0$ as functions of splitting  parameter $x_0$. For $x_0=0$ spectrum of the standard harmonic confinement  is restored. In the limit $x_0\rightarrow\infty$ two-fold degeneracy of harmonic oscillator spectrum is obtained. \label{Fig1} }
\end{figure}
\begin{equation}
V(x) = \frac{m\Omega^2}{2}(|x|-x_0)^2
\end{equation}
which is controlled by one parameter $x_0$ of the dimension of length. As it is seen, for $x_0=0$ the potential has a form of standard harmonic confinement. For $x_0>0$ it is a symmetric double-well potential formed by two harmonic confinements of frequency $\Omega$ with minima shifted by $\pm x_0$. Then, the barrier of the potential between wells is equal $x_0^2/2$ (see Fig. \ref{Fig1}). One of the advantages of this double-well model is the fact that the eigenenergies $\epsilon_i$ and the eigenstates $\psi_i(x)$ of the single-particle part of the Hamiltonian are known and can be expressed by the parabolic cylinder function (known also as the Webber function) (see \cite{MerzbacherBook} for details).
The spectrum of the single-particle Hamiltonian as a function of the parameter $x_0$ is presented in Fig. \ref{Fig1}. As it is expected, for large $x_0$ a quasi-degeneracy between even and odd eigenstates is present. In the limit $x_0\rightarrow\infty$ doubly degenerate spectrum of standard harmonic oscillator is obtained. For dynamical problems in a double-well potential it is convenient to introduce basis of states localized in the left $\varphi_{Li}(x)$ and in the right $\varphi_{Ri}(x)$ well of the potential. They are simple superpositions of neighboring eigenstates of the Hamiltonian:
\begin{subequations}
\begin{eqnarray}
\varphi_{Li}(x) = \left[\psi_{2i}(x)+\psi_{2i+1}(x)\right]/\sqrt{2},  \\
\varphi_{Ri}(x) = \left[\psi_{2i}(x)-\psi_{2i+1}(x)\right]/\sqrt{2}.
\end{eqnarray}
\end{subequations}
Of course, the localized states $\left\{\varphi_{\lambda i}\right\}$ are no longer eigenstates of the single-particle Hamiltonian. However, its matrix elements in the new basis are quite simple:
\begin{subequations}
\begin{eqnarray}
\int\mathrm{d}x\, \varphi^*_{Li}(x)\,{\cal H}_0\,\varphi_{Lj}(x) &= \delta_{ij} E_i, \\
\int\mathrm{d}x\, \varphi^*_{Ri}(x)\,{\cal H}_0\,\varphi_{Rj}(x) &= \delta_{ij} E_i, \\
\int\mathrm{d}x\, \varphi^*_{Li}(x)\,{\cal H}_0\,\varphi_{Rj}(x) &= \delta_{ij} t_i, 
\end{eqnarray}
\end{subequations}
where $E_{2k} = \frac{1}{2}(\epsilon_k+\epsilon_{k+1})$ is an average energy in a given single-particle state and $t_{2k} = \frac{1}{2}(\epsilon_k-\epsilon_{k+1})$ is a tunneling between states with the same average energy localized in opposite sites. In this basis we decompose fermionic field operators
\begin{subequations}
\begin{eqnarray}
\hat{\boldsymbol{\Psi}}_{\uparrow}(x) = \sum_i \left[\varphi_{Li}(x)\hat{a}_{Li}+\varphi_{Ri}(x)\hat{a}_{Ri}\right], \\
\hat{\boldsymbol{\Psi}}_{\downarrow}(x) = \sum_i \left[\varphi_{Li}(x)\hat{b}_{Li}+\varphi_{Ri}(x)\hat{b}_{Ri}\right],
\end{eqnarray}
\end{subequations}
where for simplicity and clarity we used different letters $a$ and $b$ for fermions with different spins. It is understood that operator $\hat{a}_{Li}$ ($\hat{a}_{Ri}$) annihilates spin-up fermion in the $i$-th state localized in the left (right) well. Operators $\hat{b}_{Li}$ and $\hat{b}_{Ri}$ act similarly on spin-down fermions.

With this decompositions the Hamiltonian (\ref{Ham1}) can be rewritten as follows:
\begin{align} \label{Ham2}
\hat{\cal H} &= \sum_i E_i \left[\hat{a}^\dagger_{Li}\hat{a}_{Li}+\hat{b}^\dagger_{Li}\hat{b}_{Li}+\hat{a}^\dagger_{Ri}\hat{a}_{Ri}+\hat{b}^\dagger_{Ri}\hat{b}_{Ri}\right] \nonumber \\
&+\sum_i t_i \left[\hat{a}^\dagger_{Li}\hat{a}_{Ri}+\hat{b}^\dagger_{Li}\hat{b}_{Ri}+\text{h.c.}\right] \nonumber \\
&+\sum_{ijkl}\sum_{\boldsymbol{\lambda}}U_{ijkl}^{\boldsymbol{\lambda}}\,\hat{b}^\dagger_{\lambda_1i}
\hat{a}^\dagger_{\lambda_2j}\hat{a}_{\lambda_3k}\hat{b}_{\lambda_4l},
\end{align}
where $\boldsymbol{\lambda}=(\lambda_1,\ldots,\lambda_4)$ is an algebraic vector of ,,left-right'' indices holding the fact that all four operators come with their own left or right basis state. 
Interaction energies can be calculated directly from the shape of localized functions
\begin{equation}
U_{ijkl}^{\boldsymbol{\lambda}} =\int \mathrm{d}x\,\varphi^*_{\lambda_1i}(x)\varphi^*_{\lambda_2j}(x)\varphi_{\lambda_3k}(x)\varphi_{\lambda_4l}(x).
\end{equation}

\begin{figure}
 \centering
 \includegraphics[scale=1.05]{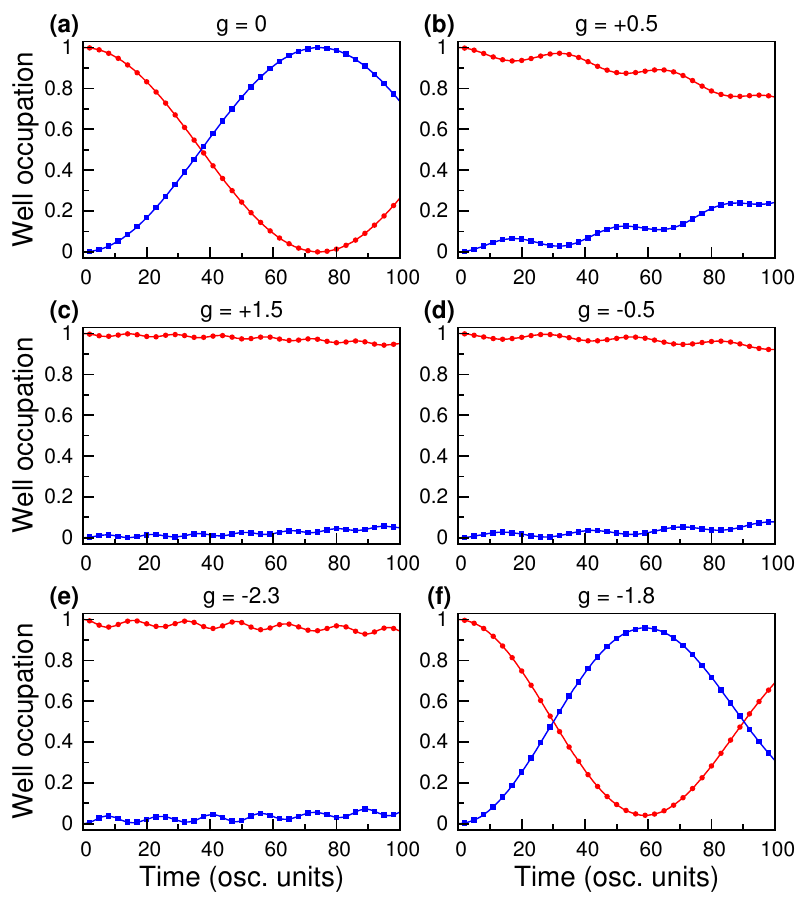}
 \caption{(color online) Occupation of spin-up (red curve) and spin-down (blue curve) particles in the left well of the confinement as functions of time for different values of interaction constant $g$. Initially the few-body state has a form \eqref{IniState2} (a) Evolution of non-interacting particles ($g=0$). In this case the evolution is governed only by single-particle tunnelings and simple Rabi oscillations are visible. (b-c) Evolution in the case of repulsive forces. Due to the energy conservation a single-particle tunneling is suppressed and oscillations are delayed. (d-e) Typical evolution for attractive interactions. As previously, a single-particle tunneling is suppressed due to misadjustment of energies between the initial state and the state with two particles occupying the same site. (f) Evolution for particular attraction $g=-1.8$ at which another few-body eigenstates of the Hamiltonian contribute to the dynamics of the initial state. An acceleration of particle flow is clearly visible. In all figures $x_0=2$ in osc. units. \label{Fig2} }
\end{figure}

\section{Dynamics of two distinguishable particles}
To get better understanding of the dynamics in the model studied, first let us concentrate on the problem of two distinguishable particles initially occupying distant wells, i.e. we assume that initially the few-body state of the system has the form
\begin{equation} \label{IniState2}
\boldsymbol{|}\mathtt{ini}\boldsymbol{\rangle} = \hat{a}_{L0}^\dagger\hat{b}^\dagger_{R0}\boldsymbol{|}\mathtt{vac}\boldsymbol{\rangle}.
\end{equation}
This particular state models a well established experimental scenario -- opposite-spin particles are prepared in distant, not coupled traps and then suddenly they are shifted closer. Due to the non-vanishing tunnelings the state is no longer an eigenstate of the Hamiltonian and it evolves in time. Note, that this preparation procedure is almost insensitive to the interaction strength $g$, since particles with opposite spins occupy distant traps. Nevertheless, interaction plays a crucial  role during the evolution. Therefore, we can study the properties of the system as a function of interaction with well established initial state. 

The evolution of the system is governed by the many-body Hamiltonian (\ref{Ham2}). Numerically, it can be determined by an exact diagonalization in the many-body basis cut at some large enough single particle level $i_{max}$. Obviously, the cut-off $i_{max}$ depends on the interaction $g$. In practice it is determined by the condition that initial state is reconstructed by the appropriate superposition of the eigenstates with the fidelity larger than $99\%$. After diagonalization, the initial state is decomposed to the eigenstates of the many-body Hamiltonian $\boldsymbol{|}\mathtt{G}_i\boldsymbol{\rangle}$ and its evolution is given explicitly
\begin{equation}
\boldsymbol{|}\Psi(t)\boldsymbol{\rangle} = \sum_i \alpha_i \mathrm{e}^{-i {\cal E}_i t/\hbar} \boldsymbol{|}\mathtt{G}_i\boldsymbol{\rangle},
\end{equation}
where $\alpha_i=\boldsymbol{\langle}\mathtt{G}_i\boldsymbol{|}\mathtt{ini}\boldsymbol{\rangle}$ is a projection of the initial state to the $i$-th eigenstate of the Hamiltonian of the energy ${\cal E}_i$. With this approach we have an access to the whole few-body state at any moment of the evolution. From this point of view the method is exact and gives complete description of the system. 
 
\begin{figure}
 \centering
 \includegraphics[scale=1.05]{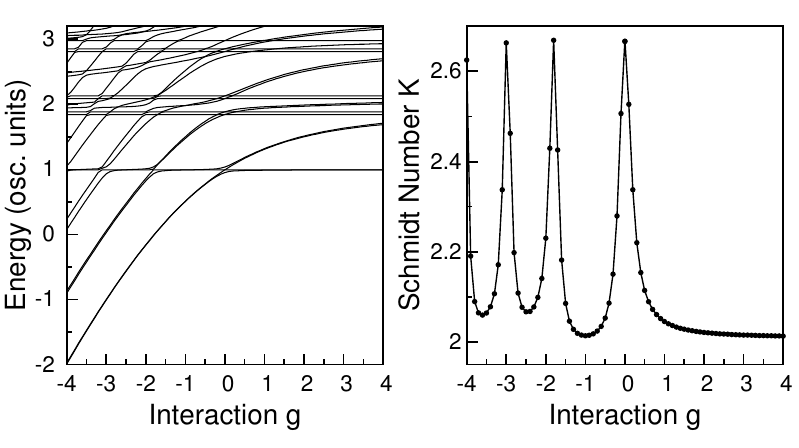}
 \caption{ (left panel) Spectrum of the many-body  Hamiltonian \eqref{Ham2} for $N_\uparrow=N_\downarrow=1$ as a function of interaction coupling $g$ near the energy of the initial state. In the limit of infinite repulsions the ground-state of the system is doubly degenerated. One of the states corresponds to the generalized Girardeau state described in the main text. For particular attractive interactions the degeneracy of the states in the vicinity of initial state energy is increased. For this interactions an acceleration of particle flow through the barrier is observed (compare with Fig. \ref{Fig2}.). (right panel) The Schmidt number $\cal K$ denoting an effective number of eigenstates of the Hamiltonian \eqref{Ham2} in the decomposition of the initial state. It is clear that $\cal K$ increases whenever acceleration in the dynamics is observed. \label{Fig3} }
\end{figure} 
 
Properties of the system during the evolution can be examined in many ways. The simplest is to study 
the evolution of the density profiles $n_\sigma(x)$ which are diagonal parts of appropriate single-particle density matrices
\begin{equation}
\rho_\sigma(x,x') = \frac{1}{N_\sigma}\boldsymbol{\langle}\Psi(t)\boldsymbol{|}\hat{\boldsymbol{\Psi}}_\sigma^\dagger(x)\hat{\boldsymbol{\Psi}}_\sigma(x')\boldsymbol{|}\Psi(t)\boldsymbol{\rangle}.
\end{equation}
Flow of the particles through the barrier can be monitored by calculating average number of particles in the left (right) well
\begin{equation}
N_{L\sigma} = \int_{-\infty}^0\!\!\mathrm{d}x\,\,n_\sigma(x), \qquad
N_{R\sigma} = \int_{0}^{+\infty}\!\!\mathrm{d}x\,\,n_\sigma(x).
\end{equation}

In the simplest case of noninteracting particles the evolution of the system is determined by the single-particle tunneling $t_0$ between the lowest single-particle levels. Simply, both particles oscillate between sites with Rabi frequency $t_0/\hbar$  (see Fig. \ref{Fig2}a).

Situation changes when repulsive interactions are present in the system. In this case higher single-particle states, due to the couplings forced by interactions, are coming into play. Typical behavior for repulsive forces are presented in Fig. \ref{Fig2}b,c. In this case mobility of particles decreases with interaction strength -- for stronger repulsions an effective tunneling is strongly suppressed. This behavior can be well explained with the energy conservation arguments. The initial state (\ref{IniState2}) has well defined energy independently of $g$ because the interaction energy, due to the almost vanishing overlap between the densities, is negligibly small. The single-particle tunneling leads to the few-body state in which both particles occupy the same potential well. In this state the interaction energy cannot be neglected and it is different than the energy of the initial state. Therefore, tunneling process is strongly suppressed due to the discrepancy of the energies. The evolution is allowed only due to the second-order process in which particles exchange the wells. An effective tunneling rate of this process is equal to $t_0^2/U$, where $U$ is the interaction energy in the intermediate state. Therefore it is suppressed by repulsions. 

Generally, the picture drawn for repulsive forces is also correct for attractions. In this case the conservation of the energy acts in the same way and strongly suppresses  the exchange dynamics (Fig. \ref{Fig2}d,e). However, for some particular attractions we find that Rabi oscillations are accelerated (Fig. \ref{Fig2}f). It happens for quite large interactions, far from the perturbative regime. Therefore, this behavior of the system cannot be explained with any single-particle picture. The explanation comes from the direct inspection of the decomposition of the initial state to the eigenstates of the Hamiltonian. The rapid oscillations occur in the system for interactions at which additional eigenstates of the Hamiltonian contribute to the state. It is indeed possible since for attractive forces the energies of excited eigenstates decrease  (left panel in Fig. \ref{Fig3}). At the same time an average energy of the initial state does not change due to the vanishing overlap mentioned above (in the case studied it is approximately equal $\hbar\Omega$). To show that this is indeed the case we calculate an effective number  of exact eigenstates ${\cal K}$ which contribute to the initial state. The number is determined in close analogy to the well known Schmidt number in the theory of reduced density matrices:
\begin{equation}
{\cal K} = \frac{1}{|\alpha_i|^4}.
\end{equation}
As it is seen in the right panel in Fig. \ref{Fig3} number ${\cal K}$ increases whenever additional eigenstates of the Hamiltonian have energy close to the energy of the initial state. The contribution of higher states to the initial state of the system has clearly resonant character. It can be viewed as an effect of resonant coupling to higher single-particle levels, i.e. the interaction term responsible for promoting two particles from the ground to the excited state of a given well is equal to the energy difference between them. If so, the single-particle tunneling in the excited level contributes to the flow of particles between sites. In consequence, particles tunnel with clearly larger frequency. From this picture it is quite obvious that in the case of two distinguishable particles this mechanism cannot be present on the repulsive side of interactions -- there is no eigenstate of the system which can contribute to the decomposition of the initial state. However, in the vicinity of $g=0$ the ground-state of the system is quasi-degenerated (as seen in Fig. \ref{Fig3}). Therefore, the peak in $\cal K$ extends into the g>0 side.

\section{Many-body dynamics of initially separated clouds}
\begin{figure}
 \centering
 \includegraphics[scale=1.05]{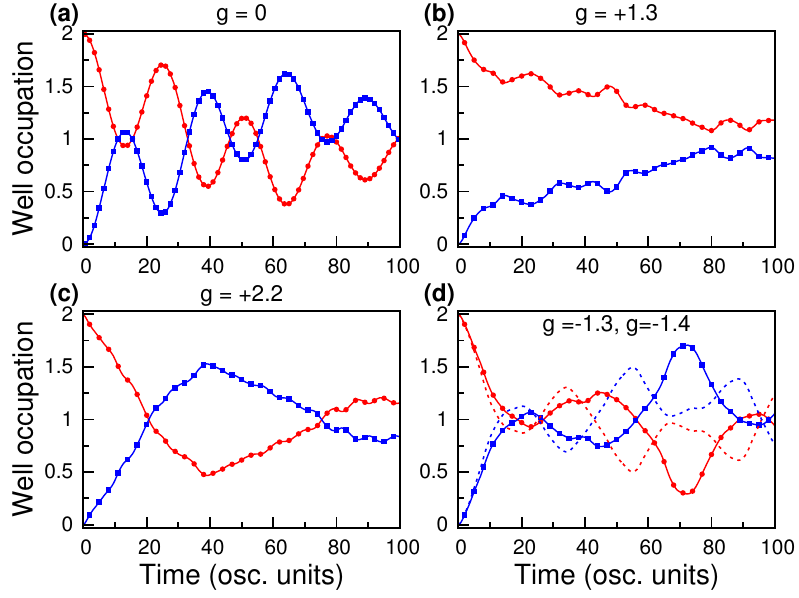}
 \caption{(color online) Similarly to Fig. \ref{Fig2}, occupation of spin-up (red curve) and spin-down (blue curve) particles in the left well of the confinement as functions of time for different values of interaction constant $g$. Here, the initial few-body state has a form \eqref{IniState4} (a) Evolution of non-interacting particles ($g=0$). In this case the evolution is governed by two different single-particle tunneling rates. Particles occupying different single-particle levels tunnel independently. In consequence, time evolution of occupations has typical two-frequency behavior.  (b-c) Evolution for two example interactions (repulsive and attractice). As previously, due to the energy conservation a single-particle tunneling is suppressed and oscillations are delayed. (d) Evolution for two, almost equal, strong attractive interactions $g=-1.3$ (solid lines) and $g=-1.4$ (dashed lines). It is clear that the dynamics in both cases is not comparable and and some kind of ,,butterfly effect'' is visible, i.e. small change in interactions leads to completely different evolution of the system. It is explained by complexity of the Hamiltonian spectrum in this region of interactions (see main text and Fig. \ref{Fig5}). In all figures $x_0=2$ in osc. units.  \label{Fig4} }
\end{figure}

Now, let us concentrate on the situation more complicated, when two pairs of particles are present.  In analogy to previous case we assume that initially both flavors are confined in distant wells of the potential, i.e. the few-body state has a form
\begin{equation} \label{IniState4}
\boldsymbol{|}\mathtt{ini}\boldsymbol{\rangle} = \hat{a}_{L0}^\dagger\hat{a}_{L1}^\dagger\hat{b}^\dagger_{R0}\hat{b}^\dagger_{R1}\boldsymbol{|}\mathtt{vac}\boldsymbol{\rangle}.
\end{equation}
We choose parameters of the potential in such a way that an overlap between particles in distant sites is still very small (here we set $x_0=2$ in osc. units). Therefore, the energy of the state is almost independent of the interaction strength $g$ and it is approximately equal to $4\hbar\Omega$. 

For non-interacting case the dynamics is governed independently at two energy levels with two different tunneling rates. In consequence, the number of particles in particular well changes with characteristic two-frequency oscillation (Fig. \ref{Fig4}a). In the presence of interactions the dynamics is qualitatively different and more complicated than in the case of two particles. For increasing repulsive interactions we find a characteristic delay in the density flow (Fig. \ref{Fig4}b). However, in the case studied, the delay is not monotonic with $g$ and a resonant behavior induced by interactions, previously observed only for attractions, is present (Fig. \ref{Fig4}c). For attractive forces the situation is even more complicated. For large enough attractions, small change of $g$ drastically changes the dynamics of the system and in practice the evolution of the densities is hard to predict. As an example in Fig. \ref{Fig4}d we plot the evolution of densities for two close interactions $g=1.3$ (red, blue lines with points) and $g=1.4$ (red, blue dashed lines). As it is seen, the similarity is visible only for short time. Later, occupations evolve completely differently. It suggests that in both cases the states evolve in completely different way. We confirm this observation also with the direct inspection of the structure of the state. Of course, since the state of the system is decomposed effectively to a finite number of eigenstates of the system there exists a finite period of time after which the system returns to the initial state. However in practice, this time is very large and many times larger than any reasonable scale of time in the problem studied.

\begin{figure}
 \centering
 \includegraphics[scale=1.05]{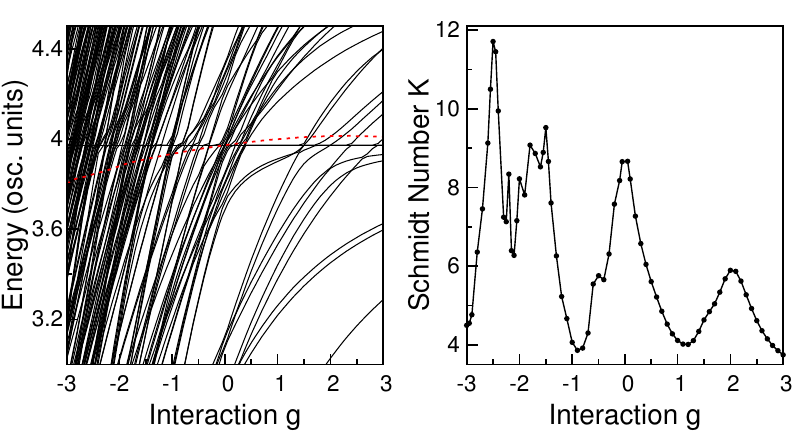}
 \caption{(color online) (left panel) Spectrum of the many-body Hamiltonian \eqref{Ham2} as a function of interaction strength $g$ for $N_\uparrow=N_\downarrow=2$. For clarity, an expectation energy of the initial state \eqref{IniState4} is marked with red dotted line. Complexity of the spectrum on the attractive side makes the decomposition of the initial state to the eigenstates of the Hamiltonian completely unpredictable. Therefore, an evolution of particles through the barrier is very sensitive to small changes of interaction $g$ (see Fig. \ref{Fig4}). This fact is also well captured by the Schmidt number $\cal K$ of effective number of eigenstates contributing to the initial state (right panel). For attractive forces $\cal K$ is very irregular function of $g$. \label{Fig5} }
\end{figure}

This highly nontrivial change in the evolution caused by a small change of interaction can be explained when the spectrum of the many-body Hamiltonian is plotted (left panel in Fig. \ref{Fig5}). For repulsive forces there are only a few eigenstates with the energies close to the energy of the initial state. Therefore, the dynamics is relatively simple and predictable. Only for some particular interactions more eigenstates contribute to the initial state and then speed-up of the transfer through the barrier is observed. For attractive forces the spectrum is essentially more complicated and a systematic description of the decomposition of the initial state to the eigenstates of the Hamiltonian is in practice impossible. This fact is reflected in a rapid changes of the Schmidt number $\cal K$ plotted in the right panel in Fig. \ref{Fig5}. We checked that the same mechanism of unpredictability is present when small changes in the double-well potential are considered. Small change of $x_0$ leads to small changes of all parameters of the Hamiltonian. However, these small changes lead to a completely different evolution of the system.

At this point it worth noticing that in the spectrum of the many-body Hamiltonian \eqref{Ham2} (left panel in Fig. \ref{Fig5}) there exists the state whose eigenenergy does not depend on the interaction $g$. This is a generalization of the well known Girardeau state to the case of a double-well potential \cite{Girardeau1,Girardeau2}. This state has a form of the single Slater determinant of the lowest $N$ single-particle orbitals of the external potential, i.e. it is totally antisymmetric under exchange of position of any two fermions independently of their spin. Consequently, interaction part of the Hamiltonian vanishes in this state. 

The results obtained may suggest that in the attractive part of the interactions the dynamics has chaotic features and can be explained by chaotic properties of the spectrum of the many-body Hamiltonian. To check this possibility we have tested the nearest neighbor spacing distribution for the energy levels in the spectrum for many different interactions. In this analysis we {\sl did not} find any traces of Wigner distribution. Contrary, the spectrum is close to the Poisson distribution. Therefore, the system is most likely not chaotic, although we have access to a limited number of eigenvalues. The dynamics is complicated and unpredictable due to the large number of eigenstates of different energies rather than to the quantum chaotic features of the spectrum.

\section{Conclusions}
To conclude, in this paper we studied dynamical properties of a few ultra-cold fermions confined in a double-well potential modeled with a double oscillator shape. We have shown that for larger number of fermions the dynamics is strongly affected by details of the potential and interaction strength between fermions of opposite spin. Particularly, flow of the particles through the barrier can be rapidly accelerated or slowed down for small adjustments of interactions. We have shown that this fact is closely related to the number of eigenstates of the full many-body Hamiltonian which has energy close to the average energy of the initial state. From this point of view it is quite natural that unpredictability of the dynamics is much stronger for attractive forces where spectrum of the Hamiltonian is much more complicated and irregular. What is worth noticing, we found that also for repulsive forces there exists such range of interactions for which the flow of particles is accelerated. Of course the results presented are just the first step to understand dynamics of few-particle systems governed by collective correlations between them. One of the simplest open questions is related to analogous properties of imbalanced mixture of fermions of the same or different masses. We believe that our findings can stimulate upcoming experiments with systems of a few ultra-cold fermions.

\section{Acknowledgments}  
The authors thank Giacomo Roati for inspiring discussions and highlighting the experimental point of view. This work was supported by the (Polish) Ministry of Sciences and Higher Education, Iuventus Plus 2015-2017 Grant No. 0440/IP3/2015/73 (T.S.) and by the (Polish) National Science Center Grant No. DEC-2012/04/A/ST2/00090 (M.G. and K.R.).

\end{document}